\documentclass[epj,final]{svjour}
\usepackage{amssymb}
\usepackage{amsmath}
\usepackage{epsfig}
\begin{document}
\title{Next-to-leading order QCD corrections to\\ 
spin-dependent hadron-pair photoproduction}
\author{
C.\ Hendlmeier\inst{1} \and  
A.\ Sch{\"a}fer\inst{1} \and
M.\ Stratmann\inst{2}\\}
\institute{
\inst{1}
Institut f{\"u}r Theoretische Physik, Universit{\"a}t Regensburg,
D-93040 Regensburg, Germany\\
\inst{2} Radiation Laboratory, RIKEN, 2-1 Hirosawa, Wako,
Saitama 351-0198, Japan}
\date{}
\abstract{We compute the next-to-leading order QCD corrections to
the ``direct'' part of the spin-dependent cross section for 
hadron-pair photoproduction. 
The calculation is performed using largely analytical methods. 
We present a brief phenomenological study of our results focussing
on the $K$-factors and scale dependence of the next-to-leading order
cross sections.
This process is relevant for the extraction of the gluon polarization in
present and future spin-dependent lepton-nucleon scattering experiments.
} 
\maketitle
%
\section{Introduction and Motivation} 
%
For many years the field of QCD spin physics has been driven 
by the hugely successful experimental program of polarized deeply-inelastic
lepton-nucleon scattering (DIS).
One of the most prominent results has been the finding 
that quarks and anti-quarks summed over all flavors, $\Delta \Sigma$, 
provide only about a quarter of the nucleon's spin, contrary to naive 
expectations from quark models.
This implies that sizable contributions to the nucleon spin should come
from the polarization of gluons, $\Delta g(Q^2)$, or from orbital angular
momenta $L_{q,\bar{q},g}(Q^2)$ of partons. Here, $Q$ denotes the resolution scale
at which the nucleon is probed.

Results from fully inclusive DIS experiments are now 
supplemented by a growing amount of data from polarized 
proton-proton collisions at BNL-RHIC \cite{ref:rhic-pol}, 
but also from less inclusive measurements in 
lepton-nucleon scattering 
\cite{ref:other-pol,ref:hermes2,ref:compass,ref:compass2}. 
Determining the gluon spin contribution is the major focus of all these experiments.
The strength of RHIC is the possibility to study several different processes
over a wide kinematical range 
which are directly sensitive to gluon polarization \cite{ref:rhic-review}: 
single-inclusive prompt photon, jet, hadron, and heavy flavor production at
high transverse momentum $p_T$ or any combination of these
final-states in two-particle correlations.
The way to access $\Delta g(Q^2)$ in lepton-nucleon scattering is to
select final-states which are predominantly produced through the
photon-gluon fusion (PGF) process. Due to the relatively small 
center-of-mass system (c.m.s.) energy $\sqrt{S}$ available in 
current fixed-target experiments, such studies are limited to
charm and single- or di-hadron production at moderate $p_T$.

To reliably determine of the amount of gluon polarization $\Delta g(Q^2)$
entering the proton helicity sum rule,
it is imperative to precisely map its Bjorken-$x$ dependence first, in
order to minimize extrapolation uncertainties in
\begin{equation}
\Delta g(Q^2) \equiv \int_0^1 \Delta g(x,Q^2) dx \;\;.
\label{eq:deltag}
\end{equation}
The eventual extraction of $\Delta g(x,Q^2)$ will require consideration
of {\em all} existing data through a ``global QCD analysis'' that makes
simultaneous use of results for all probes, from $pp$ and 
from $lN$ scattering. This is the only way to effectively deconvolute the
experimental information, which in its raw form is smeared over
the fractional gluon momentum $x$ and is taken at different scales $Q$.

The basic concept that underlies the theoretical framework for
high-$p_T$ processes, and any global analysis thereof, 
is the factorization theorem. It states that
large-mo\-men\-tum transfer reactions may be factorized into long-dis\-tance
pieces that contain the desired information on the spin structure of
the nucleon in terms of universal parton densities $\Delta f(x,Q^2)$,
$f=q,\bar{q},g$, and parts that describe the short-distance, hard
interactions of the partons. The latter can be evaluated within 
perturbative QCD. Here, at least next-to-leading order (NLO) accuracy
is required for quantitative analyses to control theoretical
uncertainties.

A first such global QCD analysis of $\Delta f(x,Q^2)$ is 
now well under way \cite{ref:dssv}, 
including all recent results from the RHIC experiments 
PHENIX and STAR \cite{ref:rhic-pol}.
However, results on ha\-dron-pair production from polarized
lepton-nucleon scattering experiments \cite{ref:hermes2,ref:compass}
have to be left out due to the complete lack of NLO computations 
for this important class of processes.
Two-hadron photoproduction is also expected to play an important role
in the spin physics program at a future polarized lepton-proton collider,
which is currently under discussion \cite{ref:erhic}. 

This paper is the first step towards a full NLO description of 
hadron-pair production in longitudinally polarized lepton-nucleon collisions.
Here, we compute the NLO QCD corrections to the ``direct'' part of the 
spin-dependent cross section for two-hadron photoproduction,
\begin{equation}
l(P_l,\lambda_l)N(P_N,\lambda_N)\rightarrow 
l'(P_{l'}) H_1(P_1) H_2(P_2) X\;\;,
\label{eq:proc}
\end{equation}
i.e., where the exchanged photon is at low virtuality and interacts
as an elementary particle with one of the partons of the nucleon $N$.
The $P_i$ in (\ref{eq:proc}) are the four-momenta of the observed
leptons and hadrons, $X$ contains all the additional hadronic
activity not observed in experiment,
and the $\lambda_i$ denote the helicities of
the interacting lepton $l$ and nucleon $N$.

Of course, an immediate complication arises here, as the direct part on its
own is no longer a well-defined quantity beyond the leading order (LO)
approximation. This is due to kinematical configurations with a collinear
splitting of the photon into a $q\bar{q}$ pair which need to be factorized
into the photon structure functions appearing in the ``resolved'' part
of the cross section. This is well-known \cite{ref:klasen} and expresses
the freedom in the factorization procedure such that
only the sum of direct and resolved contributions is independent of
theoretical conventions.
Nevertheless, we will concentrate in this work on the direct part of the
polarized two-hadron production cross section which is technically
already rather involved.

This is because we perform the calculation using large\-ly analytical methods. 
Calculations of this kind were pioneered in the unpolarized case 
for photon-hadron \cite{ref:aurenche1}, photon-photon \cite{ref:aurenche2}, 
and photon-charm \cite{ref:berger} correlations quite some time ago.
In the polarized case only a calculation for double-photon production
\cite{ref:coriano} exists so far.
To keep the computations tractable, we chose to present the results in
terms of the transverse momentum and rapidity of ``hadron one'' ($H_1$), 
$P_{T,1}$ and $y_1$, respectively, and a variable \cite{ref:aurenche1,ref:aurenche2}
\begin{equation}
z_H \equiv -\frac{\vec{P}_{T,1}\cdot \vec{P}_{T,2}}{P_{T,1}^2}
\label{eq:zhadr}
\end{equation}
which contains some information about the kinematics of ``hadron two'',
$H_2$, but not including its rapidity $y_2$. Numerical evaluations of the triple
differential cross section are thus limited in that experimental cuts
on the rapidity of the second hadron $H_2$ cannot be implemented.
The introduction of $z_H$ in the analytical calculation is essential to
keep certain singular configurations at bay \cite{ref:aurenche1,ref:aurenche2}, 
for instance, the case when the two hadrons are produced collinearly,
as will be discussed in more detail below.

Despite the fact that our calculation is not complete in the sense
discussed above, we strongly believe our results to be very important,
both theoretically and phenomenologically. On the one hand, our
results will serve as a check on more versatile calculations
in the future using combined analytical and Monte Carlo techniques 
which we are pursuing at the moment \cite{ref:upcoming} along
similar lines as in the unpolarized case \cite{ref:unpol-mc}. 
These studies \cite{ref:upcoming}
will include also the spin-dependent resolved photon contributions at NLO accuracy.
On the other hand, it was demonstrated in recent a LO study 
\cite{ref:lopaper} that the direct photon part is responsible for the
main features of the experimentally relevant spin asymmetry
and its sensitivity to the
polarized gluon density at fixed-target experiments like COMPASS and
HERMES. The resolved photon part is non-negligible though, but merely leads to
a roughly constant shift of the spin asymmetries. We also believe 
that our numerical studies of the relevance of the NLO corrections 
and theoretical uncertainties due to variations of the factorization 
and renormalization scales already give a good indication of what 
to expect from a full NLO calculation in the future.

The paper is organized as follows: in Sec.~2 we present the details
of the calculation of the NLO QCD corrections to the direct part of
spin-dependent two-hadron photoproduction. Section~3 is devoted to a brief
numerical evaluation of our results, focussing on the relevance of the
NLO corrections and the residual scale uncertainties 
at fixed-target kinematics.
Section~4 contains the conclusions. In the Appendix we collect some
additional details of the calculation.

\section{Details of the Calculation}
%
\subsection{General Framework}
%
The process we want to consider in the following is the inclusive
production of a pair of hadrons $H_1\,H_2$ in collisions of longitudinally
polarized leptons and nucleons with four-momenta as specified in
Eq.~(\ref{eq:proc}). Both hadrons are required to be at high transverse
momentum.
As mentioned above, we consider only the direct part of the cross section,
where the exchanged photon interacts as an elementary particle. 

Since we want to perform the NLO calculation using largely analytical
methods, we are limited to observing a hadron $H_1$ with 
transverse momentum $P_{T,1}$ and rapidity $y_1$, together 
with hadron $H_2$ in the opposite hemisphere, its transverse 
momentum vector $\vec{P}_{T,2}$ constrained by $z_H$ defined 
in Eq.~(\ref{eq:zhadr}), but otherwise unspecified kinematics.
Assuming, as usual, factorization, we may then write the
NLO expression for the corresponding spin-dependent cross section 
as a convolution of the non-perturbative parton distribution and
fragmentation functions and the hard-scattering of the partons
\begin{eqnarray}
&&\frac{d\Delta\sigma^{H_1H_2}}{dP_{T,1}dy_1dz_H} 
\equiv
\frac{1}{2}\left[\frac{d\sigma^{\mathrm{H_1H_2}}_{++}}{dP_{T,1}dy_1dz_H}
-\frac{d\sigma^{\mathrm{H_1H_2}}_{+-}}{dP_{T,1}dy_1dz_H}\right]
\label{eq:xsecdef} \\
&& =
\frac{2P_{T,1}}{S}\sum_{i,j,k}
\int_{1-V+V W}^{1} \frac{dz_1}{z_1}
\int_{\frac{VW}{z_1}}^{1-\frac{1-V}{z_{1}}} \frac{dv}{v(1-v)}
\int_{\frac{VW}{v z_{1}}}^{1}\frac{dw}{w} 
\nonumber \\
&& 
\times \;
\int_{z_{\min}}^{z_{\max}} \frac{dz}{z} 
\Delta f_{\gamma}^l(x_l,\mu_f)\Delta f_i^N(x_N,\mu_f)
\frac{\alpha_s(\mu_r)\alpha_{em}}{s} \nonumber\\[2mm]
&&
\times \Bigg[\frac{d\Delta\hat{\sigma}^{(0)}_{\gamma i \to j k}(v)}{dv}
\delta (1-w)\delta(1-z) + \frac{\alpha_s(\mu_r)}{2\pi}
\nonumber\\[2mm]
&&
\times\;
\frac{d\Delta\hat{\sigma}^{(1)}_{\gamma i \to j k X}}
{dv dw dz}(s,v,w,\mu_f,\mu_f^{\prime},\mu_r,z)\Bigg] 
\nonumber\\[2mm]
&&
\times\;
D_j^{H_1}(z_1,\mu_f^{\prime})D_k^{H_2}(z_2,\mu_f^{\prime})\,.
\label{eq:xsecfact}
\end{eqnarray}
The subscripts ''$++$'' and ''$+-$'' in (\ref{eq:xsecdef}) denote the
settings of the helicities of the incoming lepton and nucleon.
We have introduced the standard hadronic invariants
\begin{eqnarray}
\label{eq:mandelhadr}
&S=(P_l+P_N)^2,\; T=(P_l-P_1)^2,\; U=(P_N-P_1)^2,
&\\[2mm]
\label{eq:vwhadr}
&\quad V=1+\frac{T}{S},\;W=-\frac{U}{S+T}\,,&
\end{eqnarray}
with four-momenta specified in (\ref{eq:proc}),
and their partonic counterparts
\begin{eqnarray}
\label{eq:mandelpart}
&s=(p_{\gamma}+p_i)^2,\; t=(p_{\gamma}-p_j)^2,\; u=(p_i-p_j)^2,
&\\[2mm]
&\quad v=1+\frac{t}{s},\; w=-\frac{u}{s+t}\,.&
\label{eq:vwpart}
\end{eqnarray}
Neglecting the masses of all particles one finds the following relations 
among the variables in Eqs.~(\ref{eq:mandelhadr})-(\ref{eq:vwpart})
\begin{eqnarray}
&s=x_l x_N S,\; t=\frac{x_l}{z_1}T,\; u=\frac{x_N}{z_1}U,&\\[2mm]
&x_l=\frac{V W}{v w z_1},\; x_N=\frac{1-V}{(1-v)z_1}&
\end{eqnarray}
with $x_e$ [$x_N$] the fraction of the longitudinal momentum of the 
lepton [nucleon] taken by the quasi-real photon [parton $i$].
In addition, $V$ and $W$ in (\ref{eq:vwhadr}) are determined by 
the observed hadron $H_1$ and the lepton-nucleon c.m.s.\ energy 
squared $S$:
\begin{equation}
V=1-\frac{P_{T,1}}{\sqrt{S}}e^{-y_1},\;
W=\frac{P_{T,1}^2}{SV(1-V)}\,.
\label{eq:vwexp}
\end{equation}

In (\ref{eq:xsecfact}) we have also introduced the partonic counterpart
of $z_H$, defined via
\begin{equation}
z = - \frac{\vec{p}_{T,j}\cdot \vec{p}_{T,k}}{p_{T,j}^2} = \frac{z_1}{z_2} z_H\,
\label{eq:zparton}
\end{equation}
with $p_{T,j}$ and $p_{T,k}$ the transverse momenta
of the final-state partons $i$ and $j$ producing hadrons $H_1$ and $H_2$,
respectively. $z_{1,2}$ are the momentum share that the hadrons $H_{1,2}$
inherit from its parent partons $j,k$ in the hadronization process.
The latter is modeled by non-perturbative functions
$D_{j,k}^{H_{1,2}}(z_{1,2},\mu_{f}')$ describing the collinear
fragmentation of the partons $j$ and $k$ into the
observed hadrons $H_1$ and $H_2$, respectively. 

The $\Delta f^N(x_N,\mu_f)$ in (\ref{eq:xsecfact}) are the 
spin-dependent parton distribution functions,
defined as usual by
\begin{equation}
\Delta f^N_i(x_N,\mu_f)= f_{i(+)}^N(x_N,\mu_f)-f_{i(-)}^N(x_N,\mu_f)\,.
\label{eq:pdfdef}
\end{equation}
The subscript $+$ $[-]$ in Eq.~(\ref{eq:pdfdef}) indicates that the 
parton's spin is aligned [anti-aligned] with the spin of 
the parent nucleon $N$.

As we only focus on the direct photon case in our calculation,
$\Delta f_{\gamma}^l(x_l,\mu_f)$ in (\ref{eq:xsecfact}) coincides with 
the spin-dependent Weizs\"acker-Williams equivalent photon spectrum, 
which reads \cite{ref:ww}
\begin{eqnarray}
\Delta f_{\gamma}^l(x_l,\mu_f) &=&\frac{\alpha_{em}}{2\pi}
\left[\frac{1-(1-x_l)^2}{x_l}\ln{\frac{Q_{max}^2(1-x_l)}{m_l^2x_l^2}}
\right.\nonumber\\
&+&\left.
2m_l^2x_l^2(\frac{1}{Q_{max}^2}-\frac{1-x_l}{m_l^2x_l^2})\right]\,,
\label{eq:wwspectrum}
\end{eqnarray}
with $m_l$ the mass of the lepton.
It describes the collinear emission of a photon with low virtuality $Q$,
less than some upper limit $Q_{\max}$ determined by the experimental
conditions. The non-logarithmic pieces in (\ref{eq:wwspectrum})
result in a small but non-negligible contribution in case of muons.

The sum in (\ref{eq:xsecfact}) runs over all possible partonic 
channels $\gamma i\to j k X$, with 
$d\Delta\hat{\sigma}^{(0)}_{\gamma i \to j k}$ and 
$d\Delta\hat{\sigma}^{(1)}_{\gamma i \to j k X}$
the associated LO and NLO longitudinally polarized partonic 
hard-scattering cross sections, respectively.
They are defined in complete analogy to Eq.~(\ref{eq:xsecdef}) and
have been stripped of trivial factors involving the electromagnetic
coupling $\alpha_{em}$ and the strong coupling $\alpha_s(\mu_r)$ evaluated
at renormalization scale $\mu_r$. As indicated in (\ref{eq:xsecfact}),
starting from the NLO level, the subprocess cross sections will
explicitly depend on $\mu_r$, as well as on the scales $\mu_f$ and
$\mu_f^{\prime}$ of the parton distribution and fragmentation functions
owing to the factorization of initial and final-state collinear singularities
to be discussed below. The calculation of the NLO 
$d\Delta\hat{\sigma}^{(1)}_{\gamma i \to j k X}/dvdwdz$ is the 
main purpose of the remainder of this paper.

A computation with largely analytical methods becomes feasible
\cite{ref:aurenche1,ref:aurenche2,ref:berger,ref:coriano} thanks
to the introduction of the variable $z$ (or $z_H$) describing $H_2$. 
The price to pay is a limited control of the kinematics of hadron $H_2$, 
most notably its rapidity $y_2$.
The main virtue of $z$ is that when integrating over phase-space
certain singular configurations of two partons can be easily avoided.
For instance, momenta $p_j$ parallel to $p_k$ corresponds to negative
values of $z$. In this case, the factorized expression (\ref{eq:xsecdef})
for the cross section is incomplete and contains uncanceled poles
which would require the introduction of additional non-perturbative 
functions describing the simultaneous fragmentation of a single parton into 
two hadrons. Situations where hadron $H_2$ is parallel to the direction
of the incoming photon are characterized by $z=0$.

For the phase-space integrations of the matrix elements 
we therefore restrict ourselves to  
\begin{equation}
z>z_{\min}>0\,,
\end{equation}
equivalent to the condition that the two hadrons are 
produced by partons in opposite hemispheres.
The final expressions for the subprocess cross sections will contain 
mathematical distributions in
$z$, i.e., $\delta(1-z)$, $1/(1-z)_+$, etc., in addition to 
similar functions in the variable $w$, which are already present in the
NLO computation of single-inclusive hadron spectra 
\cite{ref:1-incl1,ref:1-incl2}. 
The analytical results will therefore be rather involved and lengthy. 
Some technical details may be found in the Appendix.

Finally, we note that corresponding expressions for spin-averaged cross sections
are straightforwardly obtained by replacing all polarized quantities in this
subsection by their unpolarized counterparts. To the best of our knowledge,
NLO corrections have not been computed with analytical methods
in the unpolarized case either. We will therefore provide results also for
$d\hat{\sigma}^{(1)}_{\gamma i \to j k X}/dvdwdz$.

\subsection{LO Contributions}
%
As we consider only the direct part of the photoproduction of two hadrons, 
there are only two partonic channels contributing to (\ref{eq:xsecfact}) 
in the  lowest order approximation: 
photon-gluon fusion,
\begin{equation}
\label{eq:lopgf}
\gamma g \rightarrow q\bar{q}\,,
\end{equation}
and the QCD Compton process,
\begin{equation}
\label{eq:locompton}
\gamma q \rightarrow q g\,.
\end{equation}
Since at LO $s+t+u=0$ and $\vec{p}_{T,j}=-\vec{p}_{T,k}$, the partonic
cross sections are $\delta$-functions in $w$ and $z$, i.e.,
\begin{equation}
\label{eq:lodelta}
\frac{d\Delta\hat{\sigma}^{(0)}_{\gamma i \to j k}}{dv dw dz}=
\frac{d\Delta\hat{\sigma}^{(0)}_{\gamma i \to j k}}{dv}
\delta (1-w)\delta (1-z)\,
\end{equation}
as indicated in (\ref{eq:xsecfact}).
The PGF process is symmetric under exchange of $q$ and $\bar{q}$, and
the result for $\gamma q \rightarrow g q$ can be obtained by replacing $v$
by $1-v$. Explicit expressions for 
$d\Delta \hat{\sigma}^{(0)}_{\gamma i\rightarrow jk}$ can
be found, e.g., in \cite{ref:1-incl1}. 
Phenomenological studies based on LO results have been performed, for instance,
 in \cite{ref:lo-old} and, most recently, in \cite{ref:lopaper}.
The unpolarized counterparts
$d\hat{\sigma}^{(0)}_{\gamma i\rightarrow jk}$ are given, e.g., 
in \cite{ref:gordon}. 

\subsection{Computation of NLO Corrections}
%
At NLO, three different types of contributions have to
be considered and evaluated:
\begin{enumerate}
\item the interference of the tree-level amplitudes for 
the processes (\ref{eq:lopgf}) and (\ref{eq:locompton}) 
and the virtual, one-loop corrections to them;
\item the real gluon emission corrections to the tree-level processes,
i.e., $\gamma q\to q g g$ and $\gamma g \to q \bar{q} g$;
\item genuine NLO processes, i.e., 
$\gamma q \to q^{\prime} \bar{q}^{\prime} q$, $\gamma q \to q \bar{q} q$.
\end{enumerate}

To account for singularities one encounters when calculating the 
loop diagrams or when performing the phase-space integrations 
for the unobserved parton,
we use dimensional regularization, where space-time is extended to
$n=4-2\varepsilon$ dimensions. To project onto definite helicity states
for the incoming parton and photon, we adopt the standard HVBM prescription
\cite{ref:hvbm} to define $\gamma_5$ and the 
Levi-Civita tensor in $n$ dimensions.
The relevant $n$-dimensional partonic hard-scattering matrix 
elements {\em before} integration over phase-space are the same as 
for single-inclusive particle or jet production and hence well known. 

The virtual corrections can be found, e.g., in \cite{ref:1-incl1} 
and \cite{ref:gordon} in the polarized
and unpolarized case, respectively. Since they resemble the two-body 
final-state of the LO result in (\ref{eq:lodelta}), their contribution
to the ${\cal{O}}(\alpha_{em}\alpha_s^2)$ corrections is 
proportional to $\delta (1-w) \delta(1-z)$.
The polarized and unpolarized NLO matrix elements in $n$ dimensions 
with a three-parton final-state 
can be taken from Refs.~\cite{ref:1-incl1,ref:gordon} as well. 
Integrating them analytically over the phase-space of the
unobserved parton is, however, much more involved than for 
single-inclusive hadron production and special care has to be taken.

\subsection{Phase-Space Integration}
%
We consider a generic ``$2\rightarrow 3$'' photoproduction process 
\begin{equation}
\label{eq:23generic}
\gamma(p_{\gamma}) i(p_i) \rightarrow j(p_j) k(p_k) l(p_l)
\end{equation}
contributing to $d(\Delta)\hat{\sigma}^{(1)}_{\gamma i\rightarrow jkX}$ in
Eq.~(\ref{eq:xsecfact}). Partons $j$ and $k$ shall produce the two observed
hadrons in the fragmentation process. Parton $l$ remains unobserved 
and hence has to be integrated over the entire phase-space.
The phase-space of one of the observed partons, say $k$, is
constrained by $z$ but otherwise integrated. We shall perform all integrations
analytically and express the result in terms of the 
variables $v$, $w$, and $z$.

Starting from the definition of the three-particle phase-space in $n$ dimensions
\begin{eqnarray}
\nonumber
dPS_3&=&\int
\frac{d^np_j}{(2\pi)^{n-1}}\frac{d^np_k}{(2\pi)^{n-1}}\frac{d^np_l}{(2\pi)^{n-1}}
\delta(p_j^2)\delta(p_k^2)\delta(p_l^2)\\
&& (2\pi)^n \delta^{(n)}(p_{\gamma}+p_i-p_j-p_k-p_l)
\label{eq:ps3} 
\end{eqnarray}
one proceeds at first along the same steps as for a one-particle 
inclusive final-state \cite{ref:ellis}, arriving at the well-known result
\begin{eqnarray}
\nonumber
dPS_3 = \frac{s}{(4\pi)^4 \Gamma(1-2\varepsilon)} 
\left[ \frac{4\pi}{s} \right]^{2\varepsilon}
v^{1-2\varepsilon} (1-v)^{-\varepsilon} dv\\
\times  [w(1-w)]^{-\varepsilon} dw \int d\theta_1 d\theta_2
(\sin\theta_1)^{1-2\varepsilon} (\sin \theta_2)^{-2\varepsilon}\,.
\label{eq:ps3help}
\end{eqnarray}
To proceed,
we parametrize the momenta in (\ref{eq:23generic}) in the c.m.s.\ frame of
partons $k$ and $l$,
\begin{eqnarray}
\nonumber
p_k &=& \frac{\sqrt{s_{kl}}}{2} (1,p_x,\sin\theta_1\cos\theta_2,
\cos\theta_1,\hat{p}_k)\\
p_l &=& \frac{\sqrt{s_{kl}}}{2} (1,-p_x,-\sin\theta_1\cos\theta_2,
-\cos\theta_1,-\hat{p}_k)\,,
\label{eq:pkpl}
\end{eqnarray}
where $s_{kl}=(p_k+p_l)^2=sv(1-w)$. $\hat{p}_k$ denotes the $(n-4)$-dimensional
components of $p_k$ and $p_x$ is arbitrary. 
The other three momenta can be chosen to have non-vanishing spatial components
only in the $y$- and $z$-directions, and explicit parametrizations can be found in
the Appendix.
The variable $z$ in (\ref{eq:zparton}) is introduced as
\begin{equation}
z\equiv m\cdot p_k
\end{equation}
by defining an auxiliary space-like vector \cite{ref:aurenche1,ref:aurenche2}
\begin{equation}
m \equiv \frac{p_{\gamma} u + p_i t + p_j s}{tu}
\label{eq:auxvec}
\end{equation}
and using the identity
\begin{equation}
1 = \int dz \delta(z - m \cdot p_k)
\end{equation} 
to perform the integration over $\theta_1$ in (\ref{eq:ps3help}).
One finally arrives at
\begin{eqnarray}
\nonumber
\!\!\!\!\!\!\!\!\!&& dPS_3 = \frac{s}{(4\pi)^4 \Gamma(1-2\varepsilon)} 
\left[ \frac{4\pi}{s} \right]^{2\varepsilon}
v^{1-2\varepsilon} (1-v)^{-\varepsilon} dv\\
\nonumber
\!\!\!\!\!\!\!\!\!&& \times  [w(1-w)]^{-\varepsilon} \,dw\, dz\, 
2 \sqrt{\frac{w(1-v)}{1-vw}} \\
\!\!\!\!\!\!\!\!\!&& \times 
\left[\frac{1-w+4w(1-v)z(1-z)}{1-vw}\right]^{-\varepsilon}
\int d\theta_2 \sin^{-2\varepsilon} \theta_2\,,
\label{eq:ps3unp}
\end{eqnarray}
in agreement with the result given in \cite{ref:aurenche1,ref:aurenche2}.
Further integration over $\theta_2$ depends on the various combinations
of scalar products of parton's momenta appearing in the hard-scattering 
matrix elements. Structures with a complicated dependence on $\theta_2$
have to be decomposed into a basic set of calculable integrals using momentum
conservation and extensive partial fractioning.
An extensive list of basic integrals can be found in 
\cite{ref:aurenche1,ref:aurenche2,ref:berger,ref:coriano} and need not be
repeated here. We only note that after integration over $\theta_2$ 
one ends up with ``plus-distributions'' both in $w$ {\em and} in $z$. We refer 
to the Appendix and Refs.~\cite{ref:aurenche1,ref:aurenche2,ref:berger,ref:coriano}
for a more detailed discussion.

Equation (\ref{eq:ps3unp}) is sufficient for integrating the unpolarized
matrix elements and most of the terms in the longitudinally polarized ones.
In the latter case, an extra complication arises, however, from 
contributions proportional to the $(n-4)$-dimensional components 
$\hat{p}_k$ in (\ref{eq:pkpl}), the so-called ``hat-momenta''.
We encounter terms in the polarized matrix elements which
depending on $\hat{p}_k^2$ and require modifications to (\ref{eq:ps3unp}),
since in its derivation we have assumed that the $(n-4)$-dimensional part 
can be trivially integrated.
Upon a careful re-examination of the steps leading to (\ref{eq:ps3unp}),
we find that these contributions lead to a modified phase-space formula 
given by
\begin{eqnarray}
\nonumber
\!\!\!\!\!\!\!\!\!&&d\widehat{PS}_3 = (-\varepsilon) 
\frac{s^2}{(4\pi)^4 \Gamma(2-2\varepsilon)} 
\left[ \frac{4\pi}{s} \right]^{2\varepsilon}
v^{2-2\varepsilon} (1-v)^{-\varepsilon} dv\\
\nonumber
\!\!\!\!\!\!\!\!\!&& \times  (1-w) [w(1-w)]^{-\varepsilon} \,dw\, dz\, 
\sqrt{\frac{w(1-v)}{1-vw}} \\
\!\!\!\!\!\!\!\!\!&& \times 
\left[\frac{1-w+4w(1-v)z(1-z)}{1-vw}\right]^{1-\varepsilon}
\int d\theta_2 \sin^{2-2\varepsilon} \theta_2\,,
\label{eq:ps3pol}
\end{eqnarray} 
in agreement with the result given in Ref.~\cite{ref:coriano}\footnote{We note
that some of the equations in the Appendix of Ref.~\cite{ref:coriano} contain
obvious misprints.}.
As is explicit in (\ref{eq:ps3pol}), all contributions stemming 
from $\hat{p}_k^2$ are of ${\cal{O}}(\varepsilon)$ as they should.
Nevertheless, they can lead to finite contributions in the limit 
$\varepsilon \rightarrow 0$ whenever they pick up a $1/\varepsilon$ pole
in the remaining phase-space integrations. 

\subsection{Factorization and Final Results}
Adding the renormalized virtual corrections and the real contributions, 
all infrared singularities cancel out, including all $1/\varepsilon^2$
terms. The remaining $1/\varepsilon$ singularities are of collinear
origin and arise when the momentum $p_l$ of the unobserved parton $l$  
becomes parallel to any of the other parton momenta.
Singular configurations related to the initial-state  
are absorbed at a factorization scale $\mu_f$ into the definition of the 
parton densities. Similarly, final-state mass singularities are
factorized at a scale $\mu_f^{\prime}$ into the bare fragmentation functions.
This is the essence of the factorization theorem.
A special role play the singularities from a collinear splitting 
$\gamma\rightarrow q\bar{q}$, which are absorbed into the photon structure
functions. Due to the freedom in choosing the amount of finite pieces
subtracted along with the pole terms, only the sum of direct and resolved
photoproduction cross sections is independent of theoretical 
conventions at NLO and beyond. 
We choose the common $\overline{\mathrm{MS}}$ scheme throughout.

As already mentioned in Sec.~2.1, the main virtue of introducing the
variable $z$ and demanding $z>0$ is to avoid certain singular 
contributions which are beyond the factorized framework outline here: (a), when
the two produced hadrons are collinear, corresponding to negative values of $z$
and (b), when hadron $H_2$ is produced parallel to the direction of the
incoming photon, characterized by $z=0$.

The factorization procedure is performed in the standard way \cite{ref:ellis}
by adding an appropriate ``counter cross section'' $d\Delta\sigma^{fact}$
to each partonic subprocess. 
At NLO, with partons $j$ and $k$ being observed as hadrons $H_1$ and $H_2$, there
are in principle four possible collinear configurations, and
$d\Delta\sigma^{fact}$ schematically reads
\begin{eqnarray}
\nonumber
&&\frac{1}{sv}\frac{d\Delta\sigma^{fact}}{dv dw dz} = \\
\nonumber 
&&-\frac{\alpha_s}{2\pi} \Bigg[
\frac{1}{sv}\Delta H_{m\gamma}[w,\mu_f] 
\frac{d\Delta\hat{\sigma}_{mi\rightarrow jk}^{\varepsilon}}{dv}[ws,v] \delta(1-z) \\
\nonumber
&&\quad\quad\quad +\frac{1}{s(1-vw)} 
\Delta H_{mi}\left[\frac{1-v}{1-v w},\mu_f \right]\\
\nonumber
&&
\quad\quad\quad\quad \times 
\frac{d\Delta \hat{\sigma}_{\gamma m \rightarrow jk}^{\varepsilon}}{dv} 
\left[\frac{1-v}{1-v w}s,vw \right] \delta(1-z)\\
\nonumber
&&\quad\quad\quad + \frac{1}{s(1-v+vw)} H_{jm}\left[1-v+v w,{\mu_f^{\prime}}\right] \\
\nonumber
&&\quad\quad\quad\quad \times 
\frac{d\Delta\hat{\sigma}_{\gamma i\rightarrow mk}^{\varepsilon}}{dv}
\left[s,\frac{vw}{1-v+v w}\right] \delta(z_1-z) \\
\nonumber
&&\quad\quad\quad + \frac{1}{sv}
H_{km}[z,{\mu_f^{\prime}}]
\frac{d\Delta\hat{\sigma}_{\gamma i\rightarrow jm}^{\varepsilon}}{dv}[s,v]\\
&&\quad\quad\quad\quad \times \theta(1-z)\delta(1-w)
\Bigg]\,.
\label{eq:fact}
\end{eqnarray}
The $d\Delta\hat{\sigma}_{ab \rightarrow cd}^{\varepsilon} [\zeta s,\xi] /dv$
are the $n$-dimensional $2\rightarrow 2$ 
cross sections for the process $ab\rightarrow cd$ 
to be found in the HVBM scheme in \cite{ref:gv}. 
These cross sections are evaluated at
some shifted kinematics denoted by $[\zeta s,\,\xi]$, since the collinear parton $j$
takes away a certain fraction of the available momentum.
Furthermore,
\begin{equation}
(\Delta)H_{ab}(\kappa,\mu)=-\frac{1}{\hat{\varepsilon}}
(\Delta)P_{ab}(\kappa)\left( \frac{\mu^2}{Q^2}\right)^{\varepsilon} + 
(\Delta)h_{ab}(\kappa)\,,
\label{eq:transfct}
\end{equation}
where $1/\hat{\varepsilon}=1/\varepsilon-\gamma_E+\ln{4\pi}$ in the 
$\overline{\mathrm{MS}}$ scheme and $z_1 \equiv 1/(1-v+v w)$.
In (\ref{eq:transfct}) the $(\Delta)P_{ab}(z)$ denote the usual 
unpolarized (polarized) one-loop splitting functions in four dimensions. 
Note that the unpolarized $H_{ab}$ contributes to the factorization of 
final-state singularities since we do not consider the production of
polarized hadrons.
The functions $(\Delta)h_{ab}(\kappa)$ represent the freedom in choosing a 
factorization prescription, and they all vanish in the $\overline{\mathrm{MS}}$ 
scheme, except for $\Delta h_{qq}(\kappa)=-16(1-\kappa)/3$ \cite{ref:trafo}.
Needless to say that an equation similar to (\ref{eq:transfct}) holds
in the unpolarized case.

For all subprocesses, the final polarized (and unpolarized)
partonic cross sections at NLO accuracy in (\ref{eq:xsecfact}) 
can be schematically cast into the following form
\begin{eqnarray}
\frac{d\Delta\hat{\sigma}^{(1)}_{\gamma i\rightarrow jkX}}{dvdwdz}
&=& \Delta K_1(v,w)\delta(1-z)
 + \Delta K_2(v,w)\delta(z-z_1) 
\nonumber\\
&+& \Delta K_3(v,w)\frac{\theta (1-z)}{(1-z)_+}+
\Delta K_4(v,w)\frac{\theta (z_1-z)}{(z_1-z)_+}
\nonumber\\
&+& \Delta K_5(v,w)\frac{\theta (z-1)}{(z-1)_+}+
\Delta K_6(v,w)\frac{\theta (z-z_1)}{(z-z_1)_+}
\nonumber\\
&+& \Delta K_7(v,w)\left(\frac{\ln (1-z)}{1-z}\right)_+ 
\nonumber\\
&+& \Delta K_8(v,w,z)\,.
\label{eq:kcoeff}
\end{eqnarray}
The coefficients $\Delta K_i$, $i=1,\ldots,8$, 
contain, in general, distributions in $w$, 
and they can be decomposed further as\footnote{The coefficients $(\Delta) K_i$ are
too lengthy to be given here, but are available upon request from the authors.} 
\begin{eqnarray}
\!\!\!\!\!\Delta K_i(v,w) &=&
\Delta k_1(v) \delta(1-w) + \Delta k_2(v)\frac{1}{(1-w)_+}
\nonumber\\
&+& \Delta k_3(v)\left(\frac{\ln(1-w)}{1-w}\right)_+
+\Delta k_4(v,w)\,.
\label{eq:kcoeffdecomp}
\end{eqnarray}
For brevity we have suppressed any dependence on the 
renormalization and factorization scales in (\ref{eq:kcoeff}) 
and (\ref{eq:kcoeffdecomp}).
Again, expressions similar to (\ref{eq:kcoeff}) and (\ref{eq:kcoeffdecomp})
hold for each unpolarized NLO partonic subprocess.

\section{Phenomenological Results}
%
We now turn to a brief numerical study of our results focussing on
the relevance of the NLO corrections and the residual scale uncertainties
for both the polarized and unpolarized cross section.
We postpone a detailed phenomenological study to a future publication
\cite{ref:upcoming} until the resolved photon contribution becomes available as well.
For our studies here, we choose the kinematical setup of the COMPASS 
experiment, which scatters a beam of polarized muons with an 
energy of $E_{\mu}=160\,\mathrm{GeV}$ off deuteron 
in a polarized $^6$LiD solid-state target, 
corresponding to a lepton-nucleon c.m.s.\ energy 
of $\sqrt{S}\simeq 18\,\mathrm{GeV}$. 

The results we show will be differential in the transverse momentum
$P_{T,1}$ of hadron $H_1$ and integrated over the angular acceptance
of the COMPASS experiment, i.e., covering scattering angles
of less than $180\,\mathrm{mrad}$. Using $y=-\ln\tan (\theta/2)$ this
straightforwardly translates into a lower bound on the pseudo-rapidity
$y_1$ for hadron $H_1$. 
Kinematics dictates the upper bound, depending on the hadron's
transverse momentum $P_{T,1}$. 
Recall that we can not control the rapidity of hadron
$H_2$ in our analytical calculation, which in turn implies that it may
end up outside the acceptance of COMPASS. The range of the transverse
momentum vector $\vec{P}_{T,2}$ of $H_2$ is restricted by demanding $z_H>0.4$,
with $z_H$ defined in Eq.~(\ref{eq:zhadr}). 
The momentum distribution of the quasi-real photons radiated off the 
muons is described by the Weizs\"{a}cker-Williams spectrum given
in (\ref{eq:wwspectrum}), with $m_l=m_{\mu}$ and $Q_{\max}^2=0.5\,\mathrm{GeV}^2$.
The photon's momentum fraction $x_l$ is restricted to be in the range
$0.1\le x_l\le 0.9$.

In the computation of the LO and NLO unpolarized cross section we use the
LO and NLO CTEQ6 parton densities \cite{ref:cteq} 
and strong coupling $\alpha_s$, respectively.
In the polarized case, we use a special set of the GRSV helicity-dependent
parton densities \cite{ref:grsv}, characterized by a small negative 
total gluon polarization
of $\Delta g=-0.15$ at the low input scale of GRSV. A small gluon polarization,
either positive or negative, is indicated by all presently available 
data sensitive to $\Delta g(x,\mu_f)$ 
\cite{ref:rhic-pol,ref:other-pol,ref:hermes2,ref:compass}, 
at least in the range of momentum
fractions predominantly probed by these experiments, 
which roughly amounts to $0.05\lesssim x \lesssim 0.2$.
We note that for a set with a small positive gluon polarization, like 
the ``standard scenario'' of GRSV \cite{ref:grsv}, 
which is also in agreement with current data,
one encounters strong cancellations between the
contributions from PGF and the Compton process, leading to sign changes in the
polarized cross section. This makes it rather awkward to display the results 
for the NLO corrections and the scale dependence we are interested in here. 
Hence, for our purposes we resort to the choice of $\Delta g=-0.15$.  
In the forthcoming publication \cite{ref:upcoming}, 
we will discuss in detail the sensitivity of two-hadron photoproduction 
to $\Delta g(x,\mu_f)$. 

To model the hadronization of partons $j$ and $k$ into the observed hadrons
$H_1$ and $H_2$, we use the novel set of fragmentation functions of DSS
\cite{ref:dss}.
This new set is based on a first global QCD analysis of inclusive hadron spectra in
electron-positron annihilation, DIS multiplicities, and hadron-hadron
scattering and known to describe ha\-dro\-ni\-zation fairly well
also in the energy range relevant for COMPASS \cite{ref:dss}. 
Since COMPASS does not identify different hadron species \cite{ref:compass}
and measures only the sum of charged hadrons,
we use the appropriate LO and NLO sets of DSS \cite{ref:dss}
for all our calculations.

\begin{figure}[thp]
\begin{center}
\vspace*{-.8cm}
\includegraphics[width=0.505\textwidth,clip=]{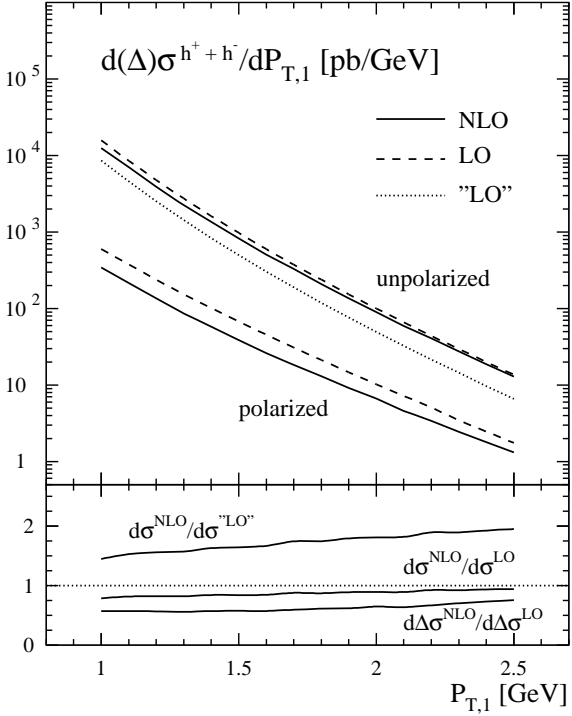}
\caption{\sf Upper panel: unpolarized and polarized photoproduction cross section
for a pair of charged hadrons, $\mu d\rightarrow (h^++h^-)(h^++h^-)X$,
at LO (dashed line) and NLO (solid line) accuracy using COMPASS kinematics.
The dotted curve labeled ``LO'' refers to a LO calculation using NLO parton
densities and fragmentation functions (see text).
The lower panel shows the corresponding
ratios of NLO to LO cross sections (K-factor).}
\label{fig:fig1}
\end{center}
\end{figure}
Figure~\ref{fig:fig1} shows our results for the $P_{T,1}$-differential
cross section for the polarized and unpolarized photoproduction of a pair
of charged hadrons at LO and NLO accuracy at COMPASS. We have set all
renormalization and factorization scales in (\ref{eq:xsecfact}) equal
to twice the transverse momentum of hadron $H_1$. The sum of the transverse
momenta of both hadrons might be a better motivated choice, but we have no control 
over $P_{T,2}$ within our analytical calculation. 
The so-called ``$K$-factor'', defined as the ratio of NLO to LO unpolarized
(polarized) cross sections
\begin{equation}
\label{eq:kfactor}
K\equiv \frac{d(\Delta) \sigma^{\mathrm{NLO}}}{d(\Delta) \sigma^{\mathrm{LO}}}\,,
\end{equation}
is depicted in the lower panel of Fig.~\ref{fig:fig1}.
The computed QCD corrections are such that the NLO results are below the LO
estimates in the entire range of $P_{T,1}$ shown in Fig.~\ref{fig:fig1}.
They appear to be more sizable in case of the polarized cross section.
The observed difference of the unpolarized and polarized 
$K$-factors clearly indicates that
NLO corrections are relevant also for studies of double-spin asymmetries,
$A_{LL}\equiv d\Delta\sigma/d\sigma$, as they do not cancel in the ratio.
The contrary is often {\em assumed} in analyses of spin asymmetries.

\begin{figure}[thp]
\begin{center}
\vspace*{-.8cm}
\includegraphics[width=0.485\textwidth,clip=]{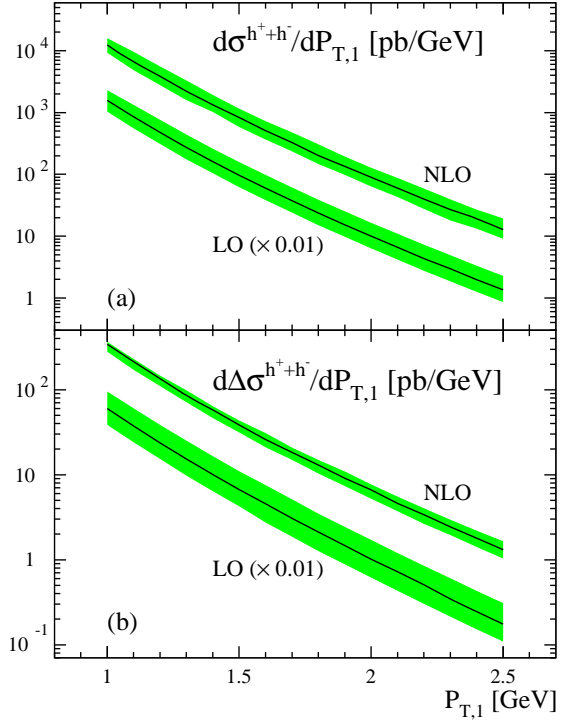}
\caption{\sf Scale dependence of the LO and NLO unpolarized 
{\bf (a)} and polarized {\bf (b)} cross sections for
$\mu d\rightarrow (h^++h^-)(h^++h^-)X$ shown in Fig.~\ref{fig:fig1}.
All scales are varied simultaneously in the range 
$\sqrt{2} P_{T,1} \le \mu_r = \mu_{f} = \mu_f^{\prime} \le 2\sqrt{2}P_{T,1}$.
Solid lines correspond to the choice where all scales are set to
$2P_{T,1}$.
All LO computations have been rescaled by a factor 0.01 to better
distinguish them from the NLO results.}
\label{fig:fig2}
\end{center}
\end{figure}
We wish to make two further remarks about the results shown in Fig.~\ref{fig:fig1}.
Firstly, finding $K$-factors smaller than one is {\em not} 
a result of the NLO corrections to the hard-scattering partonic cross sections. 
It mostly stems from the difference between
the LO and NLO parton distribution and fragmentation functions, 
in particular the latter.
The LO and NLO sets of DSS show pronounced differences, mainly because the LO
fragmentation functions try to make up for the often large NLO corrections 
in some of the fitted cross sections, see \cite{ref:dss} for details, which are
missing in a consistent LO analysis. 
The effect of the fragmentation functions is illustrated 
in Fig.~\ref{fig:fig1} in the unpolarized case by the 
curves labeled ``LO''. They refer to a calculation using LO matrix elements, but 
NLO parton densities and fragmentation functions. This clearly demonstrates the
inadequacy of LO results. At best, they can serve as a rough estimate, 
but they are insufficient for any quantitative analysis.
Very similar observations can be made in the polarized case. Here, a ``LO''-type
calculation leads to $K$-factors ranging from 1.1 to 1.8
(not shown in Fig.~\ref{fig:fig1} for clarity).
Secondly, we note that the details and size of the NLO corrections 
in the polarized case depend significantly on the still largely 
unknown gluon polarization. As already
mentioned, there can be strong cancellations between 
different subprocesses, leading
to sign changes in the polarized cross section. 
In their vicinity large NLO corrections are in general inevitable.

As an estimate for the sensitivity of the computed cross sections 
to the actual choice of scales $\mu_f$, $\mu_f^{\prime}$, 
and $\mu_r$ in (\ref{eq:xsecfact}), 
we vary them simultaneously in the range 
$\sqrt{2} P_{T,1} \le \mu_r = \mu_{f} = \mu_f^{\prime} \le 2\sqrt{2}P_{T,1}$.
We note that in principle all scales can be varied independently.
The shaded bands in Figs.~\ref{fig:fig2} (a) and (b) 
indicate the resulting residual scale
uncertainty of the unpolarized and polarized photoproduction cross sections, 
respectively.
We find that the NLO results show somewhat reduced theoretical ambiguities, 
in particular, for the polarized cross section. However, similar remarks 
as above apply also here. Scale ambiguities in the polarized case depend 
on the details of the helicity-dependent parton densities and on possible 
cancellations among different subprocesses.
One also has to keep in mind that the results so far only include the direct photon
contribution to the photoproduction cross section. It remains to be seen what the
effects of the resolved photon contribution are.
We will address all these questions in a forthcoming publication.

\section{Summary and Conclusions}
To summarize, we have computed, with largely analytical methods, the NLO
QCD corrections to the direct part of the spin-dependent cross section
for hadron-pair photoproduction.
This is the first step towards a full NLO description of this process, which
plays an important phenomenological role in the determination of the
gluon polarization in polarized lepton-nucleon collisions studied 
by HERMES and COMPASS at present and, hopefully, at higher c.m.s.\ energies
at some facility like eRHIC in the future. 

We find that the NLO corrections are essential for any study
of double-spin asymmetries. They are sizable and do not cancel in the ratio. 
Theoretical ambiguities due to the choice of the arbitrary 
renormalization and factorization scales are somewhat reduced 
if NLO corrections are taken into account.

\section*{Acknowledgements}
C.H.\ was supported by a grant of the ``Bayerische Elite\-f\"orderung''.
This work was supported in part by the ``Deut\-sche Forschungsgemeinschaft (DFG)''.

\section*{Appendix}
\subsection*{Parametrization of Momenta}
%
To perform the phase-space integration 
over the angle $\theta_2$ for all
$2\rightarrow 3$ subprocesses $\gamma i \rightarrow jkl$
analytically, we choose to work in the c.m.s.\ frame of
the observed parton $k$ and the unobserved parton $l$. 
Their momenta are parametrized in Eq.~(\ref{eq:pkpl}).
All partons are assumed to be massless.
The remaining three momenta are chosen in such a way that
they have non-vanishing components only in two spatial
directions: 
\begin{eqnarray}
p_{\gamma}&=&\frac{sv}{2\sqrt{s_{kl}}}
(1,0,\sin{\psi},\cos{\psi},\ldots)\,,\nonumber\\
p_i&=&\frac{s(1-vw)}{2\sqrt{s_{kl}}}
(1,0,-\sin{\psi},\cos{\psi},\ldots)\,,\nonumber\\
p_j&=&\frac{s(1-v+vw)}{2\sqrt{s_{kl}}}
(1,0,\sin{\psi^\prime},\cos{\psi^\prime},\ldots)\,.
\label{eq:mompara}
\end{eqnarray}
The ellipsis in (\ref{eq:mompara}) denote zeros in $(n-4)$-dimensional 
components, $s_{kl}=sv(1-w)$, and where
\begin{eqnarray}
\cos{\psi} & = & \sqrt{\frac{w(1-v)}{1-vw}}\,,\nonumber\\
\sin{\psi} & = & \sqrt{\frac{1-w}{1-vw}}\,,\nonumber\\
\cos{\psi^\prime} & = & \frac{1+v-vw}{1-v+vw}\cos{\psi}\,,\nonumber\\
\sin{\psi^\prime} & = & -\frac{1-v-vw}{1-v+vw}\sin{\psi}\,,
\end{eqnarray}
with $v$ and $w$ defined in (\ref{eq:vwpart}).

The five momenta can be used to define ten different scalar products or Mandelstam
variables. Due to momentum conservation, $p_{\gamma}+p_i=p_j+p_k+p_l$,
only five of the ten scalar products are independent. We make extensive use
of all the relations among different Mandelstam variables 
to reduce the NLO matrix elements to a form amenable to analytic integration.

In the parametrization (\ref{eq:pkpl}) and (\ref{eq:mompara}), the auxiliary
vector $m$, introduced in (\ref{eq:auxvec}), reads
\begin{equation}
m=\sqrt{\frac{s}{tu}} \left(
\sqrt{\frac{w(1-v)}{1-w}},0,0,\sqrt{\frac{1-vw}{1-w}},\ldots \right)\,.
\end{equation}

\subsection*{Plus-Distributions}
%
In the analytic calculation for two-hadron production one 
encounters not only the usual plus-distributions in the variable $w$
\cite{ref:ellis},
but also a host of distributions in $z$, the partonic counterpart of $z_H$,
as indicated in Eqs.~(\ref{eq:fact}) and (\ref{eq:kcoeff}).
For completeness, we collect here only some of the 
definitions and identities, more details can be found in 
Refs.~\cite{ref:aurenche1,ref:aurenche2,ref:berger,ref:coriano}.

With the common definition of plus-distributions via some arbitrary 
test function $f(z)$ \cite{ref:ellis}, one has
\begin{eqnarray}
\nonumber
\int_0^{z_1}dz \frac{f(z)}{(z_1-z)_+} &\equiv&
\int_0^{z_1}dz \frac{f(z_1)-f(z)}{z_1-z}\,, \\
\int_{z_1}^{z_{\max}}dz \frac{f(z)}{(z-z_1)_+} &\equiv&
\int_{z_1}^{z_{\max}}dz \frac{f(z)-f(z_1)}{z-z_1}\,,
\label{eq:zplus}
\end{eqnarray}
where $z_1=1/(1-v+vw)$ and $z_{\max}$ the upper kinematical limit 
for the $z$-integration in (\ref{eq:xsecfact}),
\begin{equation}
z_{\mathrm{max}}=\frac{1}{2}\left[1+\left(\frac{w(1-v)}{1-vw}\right)^{-1/2}
\right]\mathrm{.}
\label{eq:zbound}
\end{equation}
Analogously, the definition of other plus-distributions 
is obtained by replacing $z_1$ by $1$ in (\ref{eq:zplus}).

Note that in the case $w=1$, the upper integration limit (\ref{eq:zbound}) 
of $z$ becomes $z_{\max}=1$, and the distributions at $z=z_1$ coincide with the 
distributions at $z=1$. 
Also, in the region $z>1$, the variable $w$ is limited to $w<1$,
and distributions $1/(1-w)_+$ reduce to ordinary functions.
Finally, the lower limit $z_{\min}$ for the integration 
over $z$ in (\ref{eq:xsecfact}) is
a function of $z_H$ and the other integration variables, giving rise to additional
logarithmic contributions, e.g.,
\begin{equation}
\frac{1}{(z_1-z)_+}=\frac{1}{(z_1-z)_{z_{\min}}}+
\delta(z_1-z)\ln (z_1-z_{\min})\,,
\end{equation}
where the new distribution is defined by
\begin{equation}
\int_{z_{\min}}^{z_1}\frac{f(z)}{(z_1-z)_{z_{\min}}}dz \equiv
\int_{z_{\min}}^{z_1}\frac{f(z)-f(z_1)}{z_1-z}dz\,.
\end{equation}


%

\begin{thebibliography}{99}
%
%
\bibitem{ref:rhic-pol} PHENIX Collaboration, A.\ Adare {\em et al.},
Phys. Rev. {\bf D76}, 051106 (2007);
STAR Collaboration, B.I.\ Abelev {\em et al.},
{\tt arXiv:0710.2048 [hep-ex]}.
%
\bibitem{ref:other-pol} HERMES Collaboration, A.\ Airapetian {\em et al.}, 
Phys. Rev. Lett. {\bf 84}, 2584 (2000);
Spin Muon Collaboration (SMC), B.\ Adeva {\em et al.}, 
Phys. Rev. {\bf D70}, 012002 (2004).
%
\bibitem{ref:hermes2} P.\ Liebing (for the HERMES Collaboration), 
AIP Conf. Proc. {\bf 915}, 331 (2007).
%
\bibitem{ref:compass} COMPASS Collaboration, E.S.\ Ageev {\em et al.},
Phys. Lett. {\bf B633}, 25 (2006).
%
\bibitem{ref:compass2} COMPASS Collaboration, M.\ Alekseev {\em et al.},
{\tt arXiv:0802.3023 [hep-ex]}.
%
\bibitem{ref:rhic-review} See, for example: G.\ Bunce, N.\ Saito, J.\ Soffer, and
W.\ Vogelsang, Annu.\ Rev.\ Nucl.\ Part.\ Sci.\ {\bf 50}, 525 (2000);
C.\ Aidala et al., {\em Research Plan for Spin Physics at RHIC}, 2005, BNL report
BNL-73798-2005.
%
\bibitem{ref:dssv} D.\ de Florian, R.\ Sassot, M.\ Stratmann, and W.\ Vogelsang,
to appear.
%
\bibitem{ref:erhic} See {\tt http://www.bnl.gov/eic} for information concerning 
the eRHIC/EIC project, including the ``White Paper'' prepared for
the NSAC Long Range Plan 2007;
A.\ Deshpande, R.\ Milner, R.\ Venugopalan, and W.\ Vogelsang,
Ann. Rev. Nucl. Part. Sci. {\bf 55}, 165 (2005).
%
\bibitem{ref:klasen} See, for example: M.\ Klasen,
Rev. Mod. Phys. {\bf 74}, 1221 (2002).
%
\bibitem{ref:aurenche1} P.\ Aurenche, R.\ Baier, A.\ Douiri, M.\ Fontannaz, and
D.\ Schiff, Z.\ Phys.\ {\bf C24}, 309 (1984).
%
\bibitem{ref:aurenche2} P.\ Aurenche, R.\ Baier, A.\ Douiri, M.\ Fontannaz, and
D.\ Schiff, Z.\ Phys.\ {\bf C29}, 459 (1985).
%
\bibitem{ref:berger} E.L.\ Berger and L.E.\ Gordon, 
Phys.\ Rev.\ {\bf D54}, 2279 (1996).
%
\bibitem{ref:coriano} C.\ Coriano and L.E.\ Gordon, 
Nucl.\ Phys.\ {\bf B469}, 202 (1996).
%
\bibitem{ref:upcoming} C.\ Hendlmeier, A.\ Sch\"{a}fer, and
M.\ Stratmann, work in progress.
%
\bibitem{ref:unpol-mc} J.F.\ Owens, 
Phys. Rev. {\bf D65}, 034011 (2002);
T.\ Binoth, J.Ph.\ Guillet, E.\ Pilon, and M.\ Werlen,
Eur. Phys. J. {\bf C24}, 245 (2002).
%
\bibitem{ref:lopaper}  C.\ Hendlmeier, A.\ Sch\"{a}fer, and M.\ Stratmann,
Eur. Phys. J. {\bf C48}, 135 (2006). 
%
%
\bibitem{ref:ww} D.\ de Florian and S.\ Frixione, Phys. Lett. {\bf B457},
236 (1999).
%
\bibitem{ref:1-incl1} D.\ de Florian and W.\ Vogelsang, 
Phys. Rev. {\bf D57}, 4376 (1998);
B.\ J\"{a}ger, M.\ Stratmann, and W.\ Vogelsang, 
Phys. Rev. {\bf D68}, 114018 (2003); 
Eur. Phys. J. {\bf C44}, 533 (2005).
%
\bibitem{ref:1-incl2} B.\ J\"{a}ger, A.\ Sch\"{a}fer, 
M.\ Stratmann, and W.\ Vogelsang, Phys. Rev. {\bf D67}, 054005 (2003). 
%
%
\bibitem{ref:lo-old} M.\ Fontannaz, B.\ Pire, and D.\ Schiff,
Z. Phys. {\bf C8}, 349 (1981);
A.\ Bravar, D.\ von Harrach, and A.\ Kotzinian,
Phys. Lett. {\bf B421}, 349 (1998);
J.J.\ Peralta, A.P.\ Contogouris, B.\ Kamal, and F.\ Lebessis, 
Phys. Rev. {\bf D49}, 3148 (1994);
G.\ Grispos, A.P.\ Contogouris, and G.\ Veropoulos, 
Phys. Rev. {\bf D62}, 014023 (2000).
%
\bibitem{ref:gordon} P.\ Aurenche, R.\ Baier, A.\ Douiri, M.\ Fontannaz, and
D.\ Schiff, Nucl. Phys. {\bf B286}, 553 (1987);
L.E.\ Gordon, Phys. Rev. {\bf D50}, 6753 (1994).
%
%
\bibitem{ref:hvbm} G.\ 't Hooft and M.\ Veltman,
Nucl. Phys. {\bf B44}, 189 (1977);
P.\ Breitenlohner and D.\ Maison,
Comm. Math. Phys. {\bf 52}, 11 (1977).
%
%
\bibitem{ref:ellis} R.K.\ Ellis, M.A.\ Furman, H.E.\ Haber, and I.\ Hinchliffe,
Nucl. Phys. {\bf B173}, 397 (1980).
%
%
\bibitem{ref:gv} L.E.\ Gordon and W.\ Vogelsang,
Phys. Rev. {\bf D48}, 3136 (1993).
%
\bibitem{ref:trafo} R.\ Mertig and W.L.\ van Neerven,
Z. Phys. {\bf C70}, 637 (1996);
W.\ Vogelsang, Phys. Rev. {\bf D54}, 2023 (1996);
Nucl. Phys. {\bf B475}, 47 (1996).
%
%
\bibitem{ref:cteq} CTEQ Collaboration, 
J.\ Pumplin {\it et al.}, JHEP {\bf 0207}, 012 (2002).
%
\bibitem{ref:grsv}  M.\ Gl\"{u}ck, E.\ Reya, M.\ Stratmann, and
W.\ Vogelsang, Phys. Rev. {\bf D63}, 094005 (2001).
%
\bibitem{ref:dss} D.\ de Florian, R.\ Sassot, and M.\ Stratmann,
Phys. Rev. {\bf D75}, 114010 (2007); {\bf D76}, 074033 (2007).
%
\end{thebibliography}
\end{document}